\documentclass[twocolumn]{aastex62}
\usepackage{CJK}
\usepackage{natbib}
\newcommand{\sersic}{S\'{e}rsic}
\newcommand{\galfit}{\texttt{GALFIT}}
\makeatletter

\newcommand{\Rmnum}[1]{\expandafter\@slowromancap\romannumeral #1@}
\makeatother

\begin{document}
\begin{CJK*}{UTF8}{gkai}
\title{The Carnegie-Irvine Galaxy Survey. \Rmnum{7}. Constraints on the Origin of S0 
Galaxies from Their Photometric Structure}

\author[0000-0003-1015-5367]{Hua Gao (高桦)}
\affiliation{Department of Astronomy, School of Physics, Peking University, Beijing 100871, China}
\affiliation{Kavli Institute for Astronomy and Astrophysics, Peking University, Beijing 100871,
  China}

\author[0000-0001-6947-5846]{Luis C. Ho}
\affiliation{Kavli Institute for Astronomy and Astrophysics, Peking University, Beijing 100871,
  China}
\affiliation{Department of Astronomy, School of Physics, Peking University, Beijing 100871, China}

\author[0000-0002-3026-0562]{Aaron J. Barth}
\affiliation{Department of Physics and Astronomy, University of California at Irvine, 4129 Frederick
  Reines Hall, Irvine, CA 92697-4575, USA}

\author[0000-0001-5017-7021]{Zhao-Yu Li}
\altaffiliation{Lamost Fellow.}
\affiliation{Key Laboratory for Research in Galaxies and Cosmology, Shanghai Astronomical
  Observatory, Chinese Academy of Science, 80 Nandan Road, Shanghai 200030, China}
\affiliation{College of Astronomy and Space Sciences, University of Chinese Academy of Sciences, 19A
  Yuquan Road, Beijing 100049, China}

\begin{abstract}
  Using high-quality optical images from the Carnegie-Irvine Galaxy Survey, we perform
  multi-component decompositions of S0 galaxies to derive accurate structural parameters to
  constrain \deleted{the}\added{their} physical origin\deleted{ of this class of early-type disk
    galaxies}. Many S0s do not host prominent bulges. S0 galaxies have a broad distribution of
  bulge-to-total ratios \added{($B/T$)} and \sersic{} indices \added{($n$)}, with average values of
  $B/T=0.34\pm0.15$ and $n=2.62\pm1.02$, qualitatively consistent with the notion that S0s define a
  parallel sequence with and may have evolved from spiral galaxies. This is further reinforced by
  the incidence of bars and lenses in S0s, which when compared with the statistics in spirals,
  supports the idea that lenses are demised bars. However, despite their wide range of prominence,
  the bulges of S0s form a surprisingly \deleted{uniformly tight}\added{uniform} sequence on both
  the Kormendy and fundamental plane relations. There is no evidence for population dichotomy or
  other indications of differentiation into classical and pseudo bulges. \deleted{To the magnitude
    limit of our survey, most S0s reside in groups, and S0s in groups are as frequent as in
    clusters. Combined with previous measurements of S0 fractions in groups and clusters, we
    reemphasize that a substantial fraction of S0s have completed their morphological transformation
    before they enter clusters.}\added{Most of the S0s reside in the field and in groups;
    cluster environment is not a necessary condition for S0 production.} The properties of S0 bulges
  show little correlation with \deleted{environment}\added{environmental indicators},
  \deleted{suggesting that mass-dependent processes are more important for their
    buildup}\added{after the dependence of galaxy stellar mass on environment is taken into
    account}. As the bulges of late-type spirals and S0s are intrinsically different, and
  environmental effects that may account for such evolution appear to be minimal, we conclude that
  late-type spirals are not plausible progenitors of S0s. The bulges of S0s likely formed at an
  early epoch, after which secular processes contributed little to their subsequent evolution.
\end{abstract}

\keywords{galaxies: bulges --- galaxies: elliptical and lenticular, cD --- galaxies: evolution ---
  galaxies: photometry --- galaxies: structure}

\section{Introduction}
\label{sec:introduction}

S0 galaxies were originally proposed by \citet{1936rene+Hubble} as a hypothetical transitional class
to bridge the discontinuity between ellipticals and spirals. In later revisions and extensions of
the original Hubble classification sequence \citep[e.g.,][]{1961hag+Sandage,
  1959HDP+de_Vaucouleurs1}, S0s came to be recognized as prominent constituents of the galaxy
population. As galaxies transitioning from spirals to featureless spheroids, S0s were traditionally
thought to possess huge bulges, a preconception indeed borne out by ample empirical evidence
\citep[e.g.,][]{1979ApJ+Burstein,1985ApJS+Kent,2004ApJS+de_Souza}. Although some early observations
indicated otherwise \citep{1951ApJ+Spitzer,1970ApJ+Sandage,1976ApJ+van_den_Bergh}, the notion that
S0s can possess small bulges did not become mainstream. Only recently has the concept of a parallel
sequence of S0 galaxies \citep{1976ApJ+van_den_Bergh} resurfaced, first with the work of
\citet{2011MNRAS+Cappellari} and then extensively elaborated by \citet{2012ApJS+Kormendy}. In such a
classification scheme, S0s form a parallel sequence to spirals, one that is ordered by their bulge
prominence. S0s do not simply represent a transitional stage between spirals and ellipticals.

A major original impetus for the notion of a parallel sequence came from the realization that some
S0s have bulges as small as those in late-type spirals. The early observational evidence was
somewhat shaky because it was based on qualitative inspection of photographic plates. Systematic
investigations using modern CCD photometry have now significantly strengthened the notion that some
S0s do indeed contain small bulges \citep{2005MNRAS+Laurikainen,2006AJ+Laurikainen,
  2007MNRAS+Laurikainen,2010MNRAS+Laurikainen,2012ApJS+Kormendy}. Kinematic studies independently
reach similar conclusions \citep{2009MNRAS+Williams,2011MNRAS+Cappellari}.

Understanding the place of S0 galaxies within the framework of the Hubble sequence is not simply an
exercise in classification: it has important implications for our overall picture of
galaxy formation and evolution. The revision of the Hubble sequence to the parallel sequence
immediately suggests that S0s are red and dead spirals, which, in turn, raises the issue of what
physical processes came into play to shut off their star formation. The morphology-density relation
and its changes with redshift offer strong support for the morphological transformation of spirals
to S0s: the increase of the S0 fraction toward lower redshift counterbalances the accompanying drop
in the spiral fraction, while the elliptical population remains nearly constant
\citep[e.g.,][]{1978ApJ+Butcher,1997ApJ+Dressler, 2000ApJ+Fasano,2005ApJ+Smith}. The
morphology-density relation strongly implicates the environment as the main agent responsible for
the formation of S0s \citep[e.g.,][]{1980ApJ+Dressler, 2011MNRAS+Cappellari}. The ubiquity of lenses
and the shortage of bars in S0s compared with spirals also support an evolutionary link between
these two classes of galaxies \citep[e.g.,][]{1979ApJ+Kormendy,2009ApJ+Laurikainen,2017ApJ+Li}.

The transformation process does not lack candidate mechanisms. \citet{1951ApJ+Spitzer} first
proposed that galaxy collisions in clusters may be responsible for the loss of interstellar medium
of spirals. Later, \citet{1972ApJ+Gunn} proposed that gas stripping by the intracluster medium
produces S0s.  The fading scenario, in which spirals evolve to S0s by losing their gas via
ram-pressure stripping and then ceasing star formation, has gained growing support from a variety of
observational evidence \citep[e.g.,][]{1990AJ+Cayatte,2004AJ+Kenney,2006A&A+Aragon-Salamanca,
  2007A&A+Barr,2007ApJ+Chung,2007MNRAS+Cortese,2009AJ+Chung,2017ApJ+Poggianti} and ever more
sophisticated numerical simulations \citep[e.g.,][]{1999MNRAS+Abadi,2000Sci+Quilis,
  2007MNRAS+Roediger,2014MNRAS+Bekki,2017MNRAS+Clarke,2017MNRAS+Ruggiero}. Although ram-pressure
stripping is a viable mechanism to transform spirals into S0s in clusters, it may not be the full
story. The existence of S0s in low-density regions suggests that other factors also matter
\citep[e.g.,][]{1980ApJ+Dressler,1997ApJ+Dressler,2000ApJ+Fasano,2003MNRAS+Helsdon,2009ApJ+Wilman,
  2010ApJ+Just,2011MNRAS+Cappellari}.

At the same time, there are counterarguments against the spiral-to-S0 fading story.
\citet{2005ApJ+Burstein} found that S0s have higher $K$-band luminosities than early-type spirals,
while the fading scenario predicts the opposite. A number of studies noted that the bulges of S0s
are systematically larger than the bulges of spirals
\citep[e.g.,][]{1979ApJ+Burstein,1980ApJ+Dressler,1985ApJS+Kent,2004ApJS+de_Souza}. Although
\citet{2005MNRAS+Laurikainen,2006AJ+Laurikainen,2007MNRAS+Laurikainen,2010MNRAS+Laurikainen}
somewhat alleviated the tension by revealing that S0s, in fact, have a broad distribution of
bulge-to-total ratios, they raised other inconsistencies on the basis of the photometric structural
properties of their bulges. From another perspective, \added{the Calar Alto Legacy Integral Field
  Area (CALIFA; \citealp{2012A&A+Sanchez+CALIFA}) survey}\deleted{CALIFA observations} highlights
the discrepancy between S0s and late-type spirals in terms of their stellar angular momentum and
concentration \citep{2015A&A+Querejeta2}. Numerous alternative pathways for the evolution of S0s
have been proposed, including stellar feedback \citep{1976ApJ+Faber1}, starvation
\citep{1980ApJ+Larson,2001PASJ+Bekki,2002ApJ+Bekki,2009MNRAS+Bekki}, galaxy harassment
\citep{1996Natur+Moore, 1998ApJ+Moore,1999MNRAS+Moore}, mergers \citep{1998ApJ+Bekki,
  2015A&A+Querejeta1,2017A&A+Tapia}, and tidal interactions \citep{1985A&A+Icke,1990ApJ+Byrd,
  2011MNRAS+Bekki}. Moreover, ram-pressure stripping can still operate in groups, albeit in a more
gentle manner \citep{2006MNRAS+Rasmussen,2008ApJ+Kawata, 2008MNRAS+Rasmussen,2012ApJ+Rasmussen}. The
diversity of environments in which S0 galaxies are found may preclude us from delineating a tidy
picture of their formation. But the imprints of environments on S0s could be inferred from their
structural growth. For example, an interesting correlation between central star formation in disk
galaxies and environments was found by \citet{2004AJ+Kannappan}. Therefore, it is worthy of effort
to look into how environments might have shaped stellar structure of S0s.

In light of the aforementioned studies, we present measurements of a well-defined sample of
$z\approx 0$ S0s selected from the Carnegie-Irvine Galaxy Survey (CGS; \citealp{2011ApJS+Ho}) in
three aspects: their bulge characteristics, their incidence of bars and lenses, and their frequency
in various environments. We perform multi-component image decomposition of the sample following the
strategy outlined in \citet{2017ApJ+Gao} and confirm that S0 bulges exhibit a broad distribution of
their parameters. Comparing the bar fraction and lens fraction of CGS S0s with the statistics in
spiral galaxies suggests that lenses are demised bars.  While these observations support the picture
that S0s evolved from spirals, a major challenge is presented by the evident dearth of pseudobulges
in S0s. Various implications are discussed.

\section{Sample Definition}
\label{sec:sample-select}

The CGS sample is defined by $B_{T}\leq12.9$ mag and $\delta<0\arcdeg$, without any reference to
morphology, size, or environment. Details of the observations and data reduction are given in
\citet{2011ApJS+Ho} and \citet{2011ApJS+Li}, and will not be repeated here.  Here we focus only on
the $R$-band data. The majority of the images are of high quality, in terms of field-of-view
($8\farcm9\times8\farcm9$), median seeing ($1\farcs01$), and median surface brightness depth
(26.4\,mag~arcsec$^{-2}$).

We begin with CGS galaxies that have HyperLeda \citep{2003A&A+Paturel} morphological type index
$-3\leq T \leq 0$, complementing with CGS galaxies that fall outside this criterion but that are
classified as S0s in the Third Reference Catalogue of Bright Galaxies (RC3;
\citealp{1991Springer+de_Vaucouleurs}). We also include some possible S0s misclassified as
ellipticals, as recognized by \citet{2013ApJ+Huang} from their detailed decompositions. Finally, we
remove all galaxies with inclination angle $i>70\arcdeg$, in accordance with the selection criteria
of our training sample in \citet{2017ApJ+Gao}. The above procedure results in 94 galaxies, which
were individually decomposed following the strategy outlined in
Section~\ref{sec:decomp-strat}. Detailed examination of the decompositions compels us to remove 32
galaxies from the sample, for a variety of reasons: one galaxy is severely contaminated by too many
saturated stars; two do not belong to the formal CGS sample; three do not have available $R$-band
images; four are edge-on systems with obvious razor-thin disk, whose previously cataloged
inclination angles are clearly inaccurate; 10 are irregular/interacting galaxies or merger remnants;
six are considered ellipticals; two are actually late-type disks; and four S0s cannot be reliably
decomposed. The final sample of 62 S0s with reliable decompositions is presented in
Table~\ref{tab:bul_param}. Note that detailed fits for four of the objects (NGC~1326, 1411, 1533,
and 2784) have been presented in \citet{2017ApJ+Gao}.

\section{Decomposition Strategy}
\label{sec:decomp-strat}

Following \citet{2017ApJ+Gao}, we employ \galfit{}\footnote{
  \url{https://users.obs.carnegiescience.edu/peng/work/galfit/galfit.html}}
\citep{2002AJ+Peng,2010AJ+Peng} to perform two-dimensional (2D) multi-component decomposition of the
CGS S0s. \galfit{} is a highly flexible tool that provides many analytic functions, which allow
construction of extremely complicated models. However, we only make use of a limited set of its
features. We refer readers to \citet{2010AJ+Peng} and Section~3.2 of \citet{2017ApJ+Gao} for
detailed descriptions of the radial and azimuthal functions adopted in this study.  In brief, we
adopt the \citet{1968adga+Sersic} function for bulges, the modified Ferrer function for bars, and,
by default, the exponential function for disks, although this assumption can vary from case to
case. This study always adopts a pure ellipse for the azimuthal function. Other technical details
involved in the fitting procedure---error images, mask images, and point-spread function (PSF)
images---are described in \citet{2017ApJ+Gao}. The sky is solved simultaneously with the galaxy
model during the fitting (see Appendix~B.2 of \citealp{2017ApJ+Gao}).

\begin{figure*}[h]
  \epsscale{1.15}
  \plotone{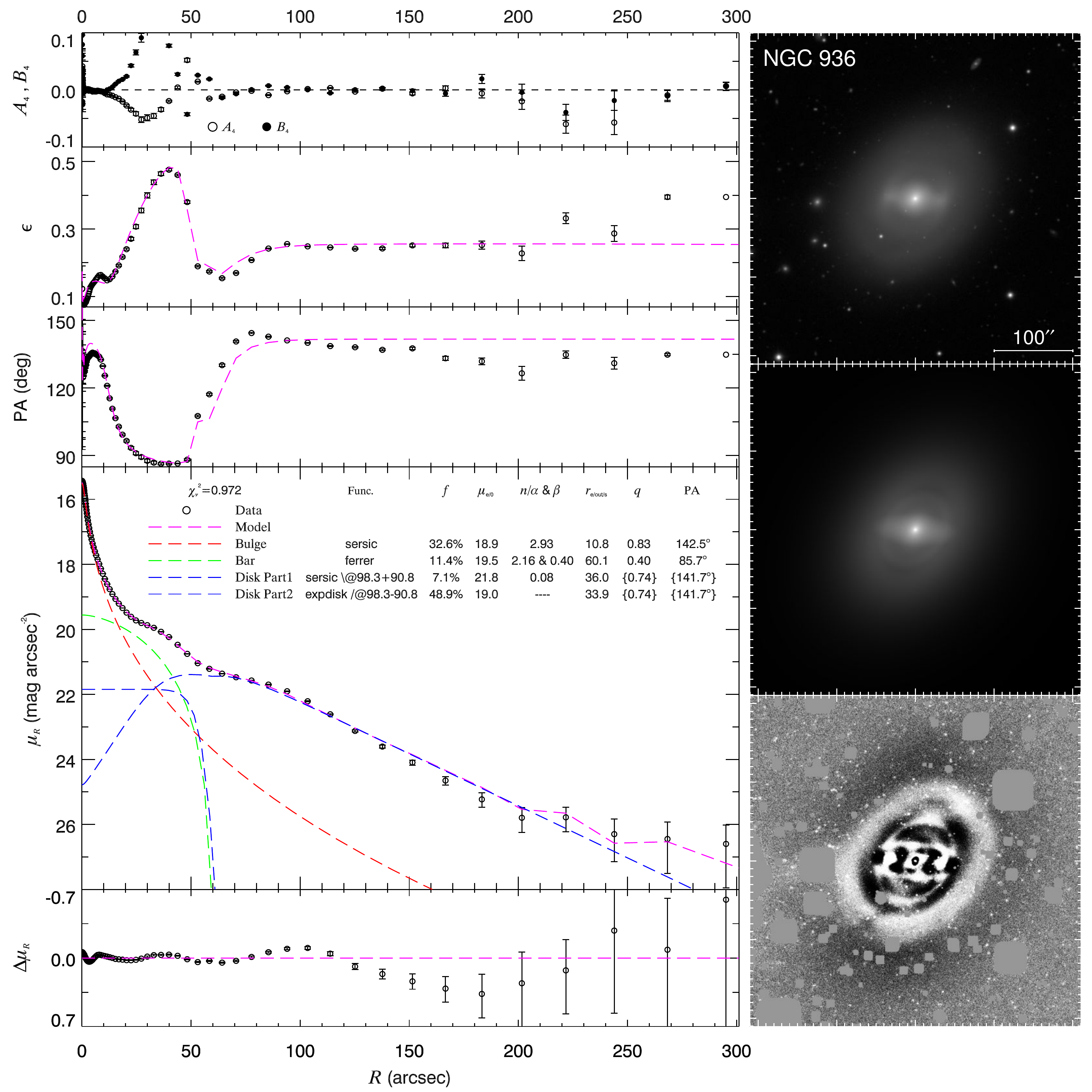}
  \caption{Best-fit model of NGC~936. The left panels display the isophotal analysis of the 2D image
    fitting. From top to bottom, the panels show radial profiles of the 4th harmonic deviations from
    ellipse ($A_{4}$ and $B_{4}$), the ellipticity ($\epsilon$), the position angle (PA), the
    $R$-band surface brightness ($\mu_{R}$), and the fitting residuals ($\bigtriangleup\mu_{R}$).
    Profiles of the data image, the model image, and the individual components are encoded
    consistently with different symbols, line styles, and colors, which are explained in the
    legends. The text on the right side of the legends gives detailed information of each component;
    from left to right, each column describes the radial profile functions (PSF, \sersic{},
    exponential, and modified Ferrer) and whether they are complete or truncated (blank for
    complete, ``\textbackslash@$r_{\mathrm{break}} +\bigtriangleup r_{\mathrm{soft}}$'' for outer
    truncation, and ``/@$r_{\mathrm{break}}-\bigtriangleup r_{\mathrm{soft}}$'' for inner
    truncation), the light fractions, the characteristic surface brightness (effective surface
    brightness $\mu_{e}$ for the bulge and central surface brightness $\mu_{0}$ for the others), the
    shape parameters of the radial profiles (\sersic{} index $n$ for the \sersic{} function and
    $\alpha\,\&\,\beta$ for the modified Ferrer function), the characteristic radii (effective
    radius $r_{e}$ for the \sersic{} function, outer boundary $r_{\mathrm{out}}$ for the modified
    Ferrer function, and scale length $r_{s}$ for the exponential function), the axis ratios ($q$),
    and the position angles (PA). The parameters can be constrained to be the same (braces) and/or
    fixed (brackets). Note that the surface brightness profile of the model is generated by fixing
    the geometric parameters to those of the data surface brightness profile, and the surface
    brightness profiles of individual components are generated along their major axes; hence, the
    model surface brightness profile is not a simple summation of those of the individual
    components. The right panels display, from top to bottom, the grayscale $R$-band data image, the
    best-fit model image, and the residual image. The images are shown using the same logarithmic
    stretch for the data and model image, and histogram equalization stretch for the residual
    image. All images are cropped to have the same size of $1.5D_{25}$ and are centered on the
    galaxy centroid, with north up and east to the left. \\(The complete figure set of 62 images is
    available in the online journal.) \label{fig:exam}}
\end{figure*}

Based on the optimal strategy for bulge decomposition investigated in \citet{2017ApJ+Gao}, we are
aware of which parts of the galaxy should be modeled or can be neglected, and of the penalties
incurred for ignoring certain parts of the galaxy in model construction. Here we prepare only
\textit{one} optimal model for each galaxy, based on recognition of its principal structural
components. In addition to bulges and disks, we model bars, disk breaks, nuclear/inner lenses, and
inner rings, but do not model nuclear rings/bars separately because we consider them as part of the
photometric bulge. Outer lenses/rings have been shown to be not crucial for measuring accurate bulge
parameters, and thus will be ignored. As S0s lack spiral arms, they, too, can be neglected. Unless
specifically mentioned in Appendix~\ref{sec:notes-indiv-galax}, we follow the above guidelines to
construct surface brightness models for our sample. The best-fit models of the 62 CGS S0s are shown
in Figure~\ref{fig:exam}, and the best-fit parameters are summarized in Table~\ref{tab:bul_param}.

The error budget of bulge parameters (total magnitude $m$, bulge-to-total ratio $B/T$, effective
surface brightness $\mu_{e}$, \sersic{} index $n$, effective radius $r_{e}$, and ellipticity
$\epsilon$) includes the uncertainties introduced by sky measurements and, more importantly, the
uncertainties introduced by model assumptions.  We measure the sky-induced uncertainties as
variations of the best-fit bulge parameters when perturbing the sky levels around \(\pm1\sigma\) of
the best-fit sky levels. An important source of model-induced uncertainty comes from ignoring
certain features of the galaxy (e.g., outer lenses and rings). Ignoring these features will induce
uncertainties of 0.05\,mag, 7.1\%, 0.09\,mag~arcsec$^{-2}$, 5.8\%, 5.3\%, and 4.8\% for $m$, $B/T$,
$\mu_{e}$, $n$, $r_{e}$, and $\epsilon$, respectively \citep{2017ApJ+Gao}. Another source of
uncertainty stems from the use of different mathematical representations of the same disk surface
brightness, which arise when we model disk breaks, lenses, and rings, along with the underlying
disk. The typical contribution to the error budget is 0.09\,mag, 6.7\%, 0.15\,mag~arcsec$^{-2}$,
8.0\%, 6.9\%, and 7.7\% for $m$, $B/T$, $\mu_{e}$, $n$, $r_{e}$, and $\epsilon$, respectively
\citep{2017ApJ+Gao}. The final uncertainties of the bulge parameters in Table~\ref{tab:bul_param}
represent the quadrature sum of these various sources of uncertainties.

\begin{deluxetable*}{l@{ }rCCCCCCDCc}
  \tabletypesize{\scriptsize}
  \tablecaption{Bulge Parameters and Bar/Lens Identifications of the CGS S0s \label{tab:bul_param}}
  \tablehead{\multicolumn2c{Name} & \colhead{$m$} & \colhead{$B/T$} & \colhead{$\mu_{e}$} &
    \colhead{$n$} & \colhead{$r_{e}$} & \colhead{$\epsilon$} & \twocolhead{Scale} &
    \colhead{$\sigma_{0}$} & \colhead{Bar/Lens} \\ \multicolumn2c{} & \colhead{(mag)} & \colhead{} &
    \colhead{(mag~arcsec$^{-2}$)} & \colhead{} & \colhead{($\arcsec$)} & \colhead{} &
    \twocolhead{($\mathrm{kpc/\arcmin}$)} & \colhead{(km~s$^{-1}$)} & \colhead{} \\ \multicolumn2c{(1)}
    & \colhead{(2)} & \colhead{(3)} & \colhead{(4)} & \colhead{(5)} & \colhead{(6)} & \colhead{(7)}
    & \twocolhead{(8)} & \colhead{(9)} & \colhead{(10)}}
  \decimals
  \startdata
  ESO & 221-G026 & 11.13\pm 0.17 & 0.534\pm 0.062 & 19.46\pm 0.46 & 5.00\pm 0.65 & 13.48\pm  3.72 & 0.528\pm 0.041 &  3.61 & 135.7\pm\phn5.1 & ? \\
  ESO & 442-G026 & 12.21\pm 0.09 & 0.341\pm 0.024 & 17.71\pm 0.15 & 1.36\pm 0.15 & \phn3.74\pm 0.26 & 0.175\pm 0.014 & 12.57 & 221.3\pm  19.4 & ? \\
  ESO & 507-G025 & 12.67\pm 0.13 & 0.185\pm 0.026 & 18.16\pm 0.17 & 1.71\pm 0.15 & \phn3.98\pm 0.33 & 0.348\pm 0.028 & 10.53 & 260.2\pm\phn9.9 & N \\
  IC &     2006 & 11.94\pm 0.43 & 0.329\pm 0.101 & 20.05\pm 0.69 & 4.02\pm 0.60 & \phn9.26\pm 3.54 & 0.100\pm 0.009 &  5.56 & 123.8\pm\phn2.3 & L \\
  IC &     2035 & 13.70\pm 0.09 & 0.109\pm 0.008 & 14.69\pm 0.15 & 0.81\pm 0.07 & \phn0.57\pm 0.04 & 0.297\pm 0.023 &  4.80 & 106.7\pm\phn2.3 & B \\
  IC &     4329 & 12.07\pm 0.06 & 0.226\pm 0.002 & 20.11\pm 0.11 & 3.22\pm 0.11 & \phn9.94\pm 0.75 & 0.182\pm 0.010 & 19.18 & 295.9\pm\phn6.0 & W \\
  \enddata

  \tablecomments{ Column 1: Galaxy name. Column 2: Total $R$-band magnitude. Column 3:
    Bulge-to-total ratio. Column 4: Surface brightness at the effective radius. Column 5: \sersic{}
    index. Column 6: Effective radius. Column 7: Ellipticity. Column 8: Scale to convert from arcmin
    to kpc. Column 9: Central stellar velocity dispersion. Column 10: Flag for the presence or
    absence of a bar/lens: \deleted{Y}\added{B} = definitely barred; W = weakly barred; N = no bar
    or lens; L = no bar but lens present; ? = uncertain. \\(Table~\ref{tab:bul_param} is published
    in its entirety in machine-readable format. A portion is shown here for guidance regarding its
    form and content.)}
\end{deluxetable*}

\section{Results}
\label{sec:results}

\subsection{The Scaling Relations of Bulges}
\label{sec:nonun-unif-s0}

\begin{figure*}
\epsscale{1.1}
\plotone{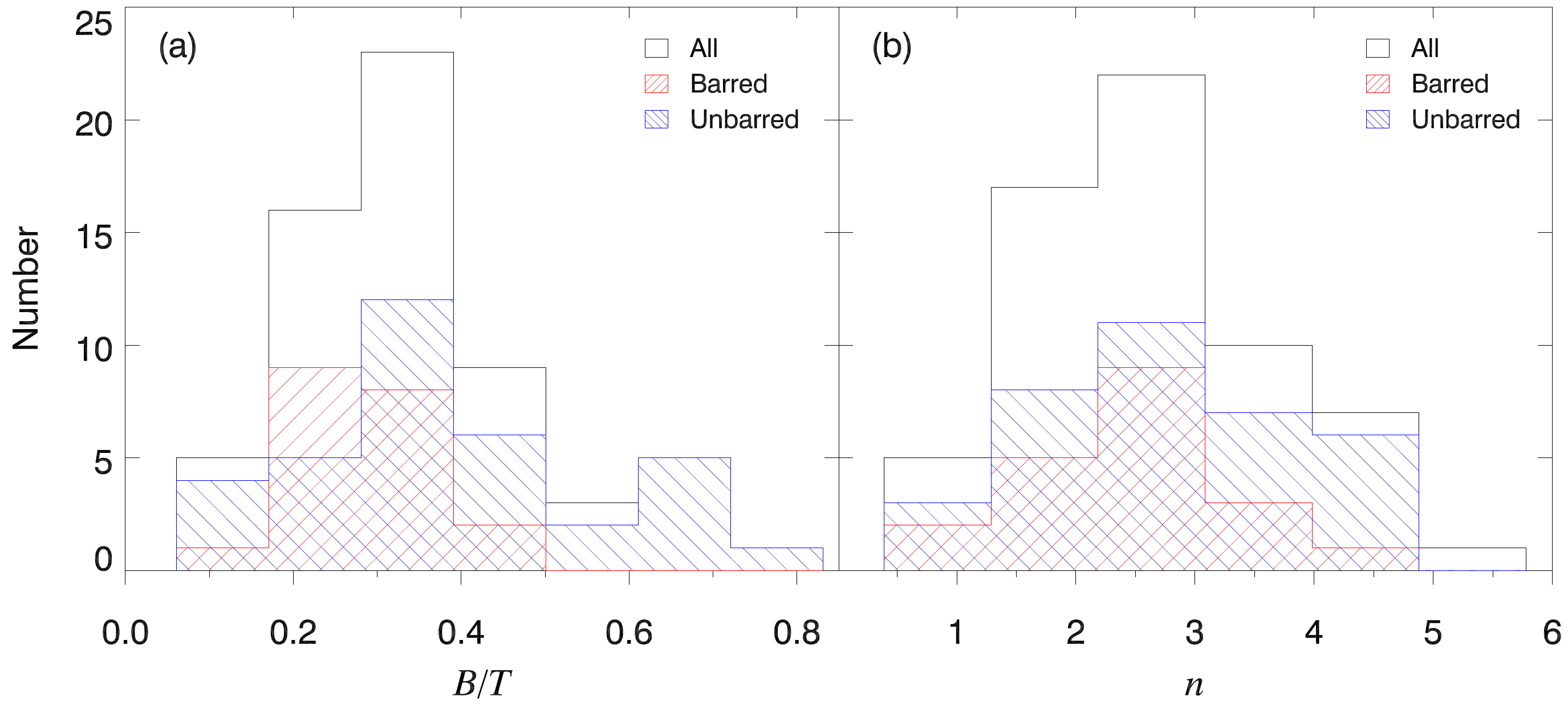}
\caption{Distributions of (a) $B/T$ and (b) \sersic{} index $n$ for all S0s (black), barred S0s
  (red), and unbarred S0s (blue). \label{fig:hist}}
\end{figure*}

As the structural parameters of the bulge ($n$, $\mu_{e}$, and $r_{e}$) are internally correlated,
we only investigate the distribution of $B/T$ and \sersic{} indices $n$ (Figure~\ref{fig:hist}). The
bulges of our sample of S0 galaxies exhibit a wide range of bulge prominence and light
concentration. Their mean $B/T$ is 0.34, with a standard deviation of 0.15; their mean $n$ is 2.62,
with a standard deviation of 1.02. The fraction of S0s that have a bulge prominence comparable to
that of typical late-type spirals, $B/T\leq0.15$, is 6.5\%. We confirm the results of the study of
\citet{2005MNRAS+Laurikainen} that S0s are less bulge-dominated than previously thought (e.g.,
$\langle B/T\rangle=0.63$ from \citealp{2004ApJS+de_Souza}; see Table~8 in
\citealp{2005MNRAS+Laurikainen} for a compilation of previous S0 measurements). We should note that
the methodology of image decomposition employed by us or by \citet{2005MNRAS+Laurikainen} differs in
detail from those of previous studies, both in terms of technique (1D vs. 2D) and choice of surface
brightness models (two-component vs. multi-component). The $B/T$ is significantly reduced when bars
and other secondary morphological features (e.g., lenses) are properly isolated from the bulge.

Figure~\ref{fig:hist} examines separately the distributions of bulge parameters for barred and
unbarred systems. The barred S0s are less bulge-dominated and show a narrower distribution of $B/T$
($\langle B/T \rangle=0.28\pm 0.07$\footnote{Note that we express statistics of distributions as
  mean$\pm$standard deviation throughout the text.}) compared with the unbarred subsample
($\langle B/T \rangle=0.38\pm 0.18$). This, again, is in line with the measurements of
\citet{2010MNRAS+Laurikainen,2013MNRAS+Laurikainen}. A Kolmogorov-Smirnov test finds that the
difference in $B/T$ between the two subsamples is statistically significant, with a probability
$P_{\mathrm{null}}=0.044$ for the null hypothesis that the two subsamples are drawn from the same
parent distribution. In term of \sersic{} indices, the differences between the two subsamples are
more subtle ($\langle n \rangle=2.55\pm0.94$ for barred S0s; $\langle n \rangle=2.70\pm1.03$ for
unbarred S0s), and not statistically significant ($P_{\mathrm{null}}=0.811$).

\begin{figure*}
\epsscale{1.18}
\plotone{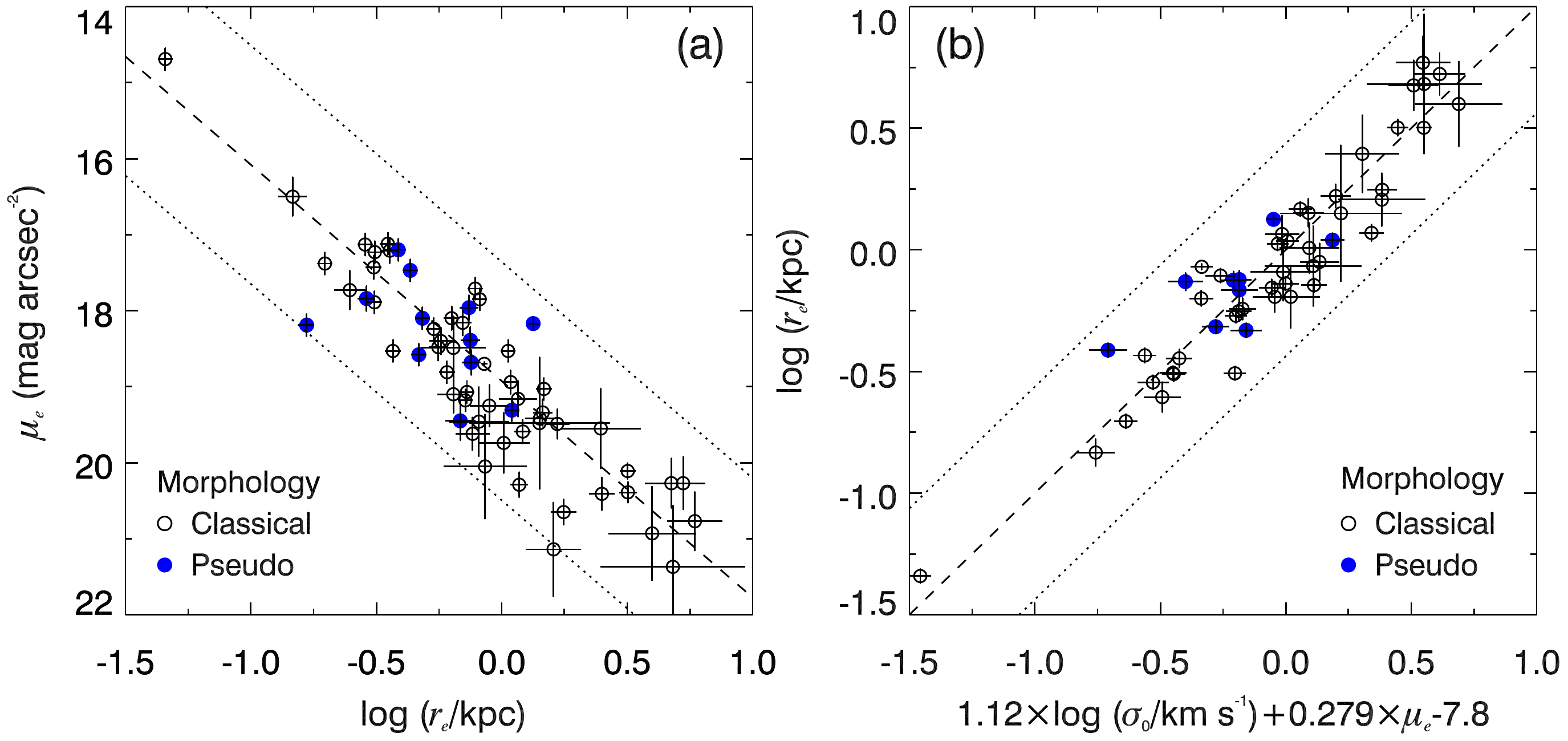}
\caption{The distribution of the bulge parameters of S0 galaxies on (a) the Kormendy relation and
  (b) the edge-on view of the fundamental plane. The dashed lines are the best-fit linear
  relations. The dotted lines mark the $3\sigma$ scatter of the best-fit relations. The blue filled
  \deleted{points}\added{circles} represent bulges with a pseudobulge appearance. \label{fig:FP}}
\end{figure*}

The distribution of structural parameters on the scaling relations of bulges are particularly useful
for distinguishing their physical nature \citep{2004ARA&A+Kormendy}. We derive the
\citet{1977ApJ+Kormendy+Kormendy_relation} relation of S0 bulges by minimizing the quantity
\begin{equation}
\label{eq:1}
\chi^{2}=\sum_{i=1}^{N}\frac{\left(\mu_{e,i}-\alpha\log r_{e,i}-
\beta\right)^{2}}{\xi_{\mu_{e,i}}^{2}+\alpha^{2}\xi_{\log r_{e,i}}^{2}},
\end{equation}
where $\alpha$ and $\beta$ are the coefficients of the Kormendy relation
$\mu_{e}=\alpha\log r_{e}+\beta$ and $\xi$ denote uncertainties. The best-fit relation is
(Figure~\ref{fig:FP}a)
\begin{equation}
\label{eq:2}
\mu_{e}=\left(2.85\pm0.07\right)\log r_{e}+\left(18.93\pm0.03\right), 
\end{equation}
with a scatter in $\mu_{e}$ of $0.52\,\mathrm{dex}$.  Similarly, we derive the fundamental plane
relation for the bulges with central stellar velocity dispersions ($\sigma_{0}$) available from
\citet{2011ApJS+Ho}, which were assembled from HyperLeda and were originally compiled from various
literature. Minimizing the quantity
\begin{equation}
\label{eq:3}
\chi^{2}=\sum_{i=1}^{N}\frac{\left(\log r_{e,i}-a\log
\sigma_{0,i}-b\mu_{e,i}-c\right)^{2}}{\xi_{\log r_{e,i}}^{2}+a^{2}\xi_{\log
\sigma_{0,i}}^{2}+b^{2}\xi_{\mu_{e,i}}^{2}},
\end{equation}
where $a$, $b$, and $c$ are coefficients of the fundamental plane relation 
$\log r_{e}=a\log \sigma_{0}+b\mu_{e}+c$, the best-fit relation is (Figure~\ref{fig:FP}b)
\begin{equation}
\label{eq:4}
\log r_{e}=\left(1.12\pm0.08\right)\log \sigma_{0}+
\left(0.279\pm0.009\right)\mu_{e}-\left(7.8\pm0.2\right).
\end{equation}
The scatter in $\log r_{e}$ is $0.15\,\mathrm{dex}$.

\begin{figure*}
\epsscale{1.18}
\plotone{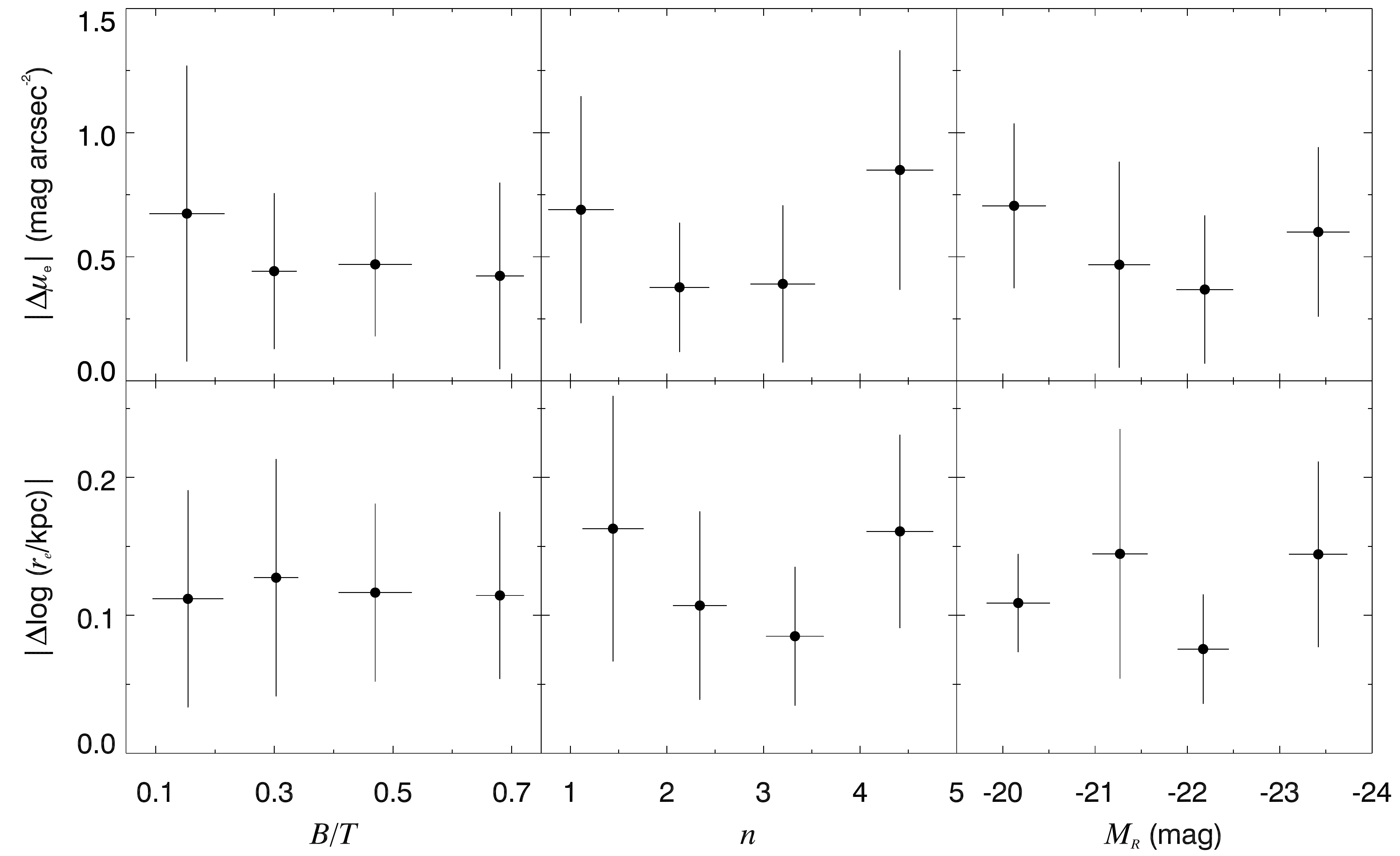}
\caption{Residuals of (top) the Kormendy relation in $\mu_{e}$ and (bottom) fundamental plane in
  $\log r_{e}$ as a function of (left) $B/T$, (middle) bulge \sersic{} index $n$, and (right)
  absolute $R$-band magnitude $M_{R}$ of the galaxy. Error bars represent standard deviation in each
  bin. There is no strong dependence on any parameter. \label{fig:FPres}}
\end{figure*}

Both scaling relations are uniformly tight across a large dynamical range in $r_{e}$. Although
pseudobulges are expected to be outliers in these scaling relations \citep[e.g.,][]{
  2004ARA&A+Kormendy,2009MNRAS+Gadotti,2010ApJ+Fisher}, and previous studies indicate that at least
some S0s host pseudobulges \citep{2005MNRAS+Laurikainen,2006AJ+Laurikainen,2007MNRAS+Laurikainen},
we are hard-pressed to identify such a population in our sample, at least on the basis of the
Kormendy and fundamental plane relations. To investigate this point further, we attempt to identify
pseudobulge candidates in our sample from careful inspection of their images, color maps, color
profiles, isophotal analysis, and residual images from the decomposition. We look for fine
structures in the bulge region suggestive of a disky nature (e.g., nuclear rings/bars) or signatures
of recent star formation (e.g., distinctly blue centers compared with their surroundings). However,
the minority of S0 bulges with pseudobulge characteristics (blue symbols in Figure~\ref{fig:FP})
appear indistinguishable from the rest of the majority.  We further look into possible dependence of
the residuals in the two scaling relations on properties considered to be related to the pseudobulge
phenomenon. Motivated by previous studies that suggest that pseudobulges have low \sersic{} indices
($n\la2$; \citealp{2008AJ+Fisher}) and small bulge-to-total ratios (Kormendy \& Ho 2013), we group
the sample into several bins of $B/T$ and $n$ and plot their residuals with respect to the best-fit
Kormendy relation and the fundamental plane (Figure~\ref{fig:FPres}).  It is obvious that there is
no clear systematic dependence of the residuals on bulge properties. We also look for systematic
trends with galaxy luminosity (absolute $R$-band magnitude $M_{R}$), but do not find any either. The
S0 bulges behave as a uniform population in the Kormendy relation and the fundamental plane
relation, despite the fact that they show broad characteristics in terms of $B/T$ and $n$.

\subsection{Bars and Lenses}
\label{sec:bar-fraction-lens}

\begin{figure}
\epsscale{1.18}
\plotone{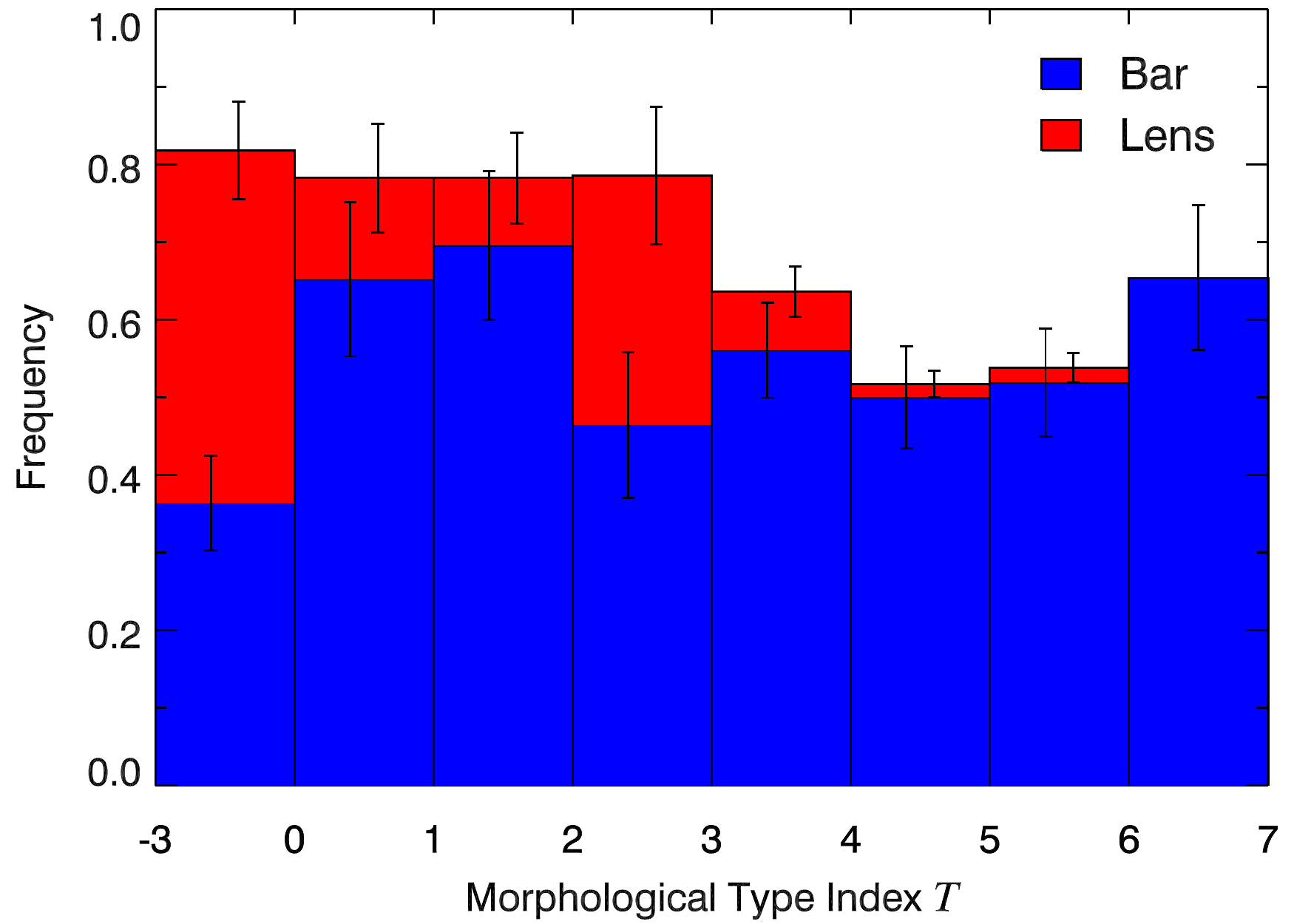}
\caption{Bar and lens frequencies for CGS disk galaxies as a function of morphological type index
  $T$. \added{Error bars are horizontally offset for clarity.} \label{fig:bar_lens_freq}}
\end{figure}

Being an exclusively disk phenomenon, bars potentially provide another link to relate S0s to
spirals. A simple fading scenario, however, probably cannot fully account for the evolution of
spirals into S0s, for S0s are known to possess fewer bars but more lenses than spirals
\citep{2009ApJ+Laurikainen}. Several lines of evidence, including their similarity in size and
stellar content, led \citet{1979ApJ+Kormendy} to propose that lenses are dissolved bars. Later
theoretical developments showed that during the course of bar-driven secular evolution gas-rich
galaxies can build up sufficiently massive central concentrations that can weaken and even dissolve
the bar \citep[e.g.,][]{1993RPPh+Sellwood,1996FCPh+Buta,1996ASPC+Combes,2004ARA&A+Kormendy,
  2004ApJ+Shen}. According to \citet{2018MNRAS+Kruk}, the host galaxies of lenses have similar
properties as those hosting bars. If lenses are indeed evolved bars, they may help to bridge the gap
between S0 and spiral statistics. To test this scenario, we measure the bar fraction and lens
fraction of the CGS S0s, and compare them with the bar fractions of CGS spirals. The identification
of bars and lenses for the S0s are given in Table~\ref{tab:bul_param}. The CGS S0 sample contains a
bar fraction of $36\pm6\%$ and a lens fraction of $45\pm6\%$. Thus, the combined fraction of S0
galaxies hosting either a bar or a lens is $81\pm5\%$. We also quantify the bar fraction of
\deleted{$\sim300$}\added{291} non-edge-on ($i\leq70\arcdeg$) CGS spirals (morphological type index
\deleted{$T=0-7$}\added{$0<T\leq7$}) by inspecting their images, isophotal properties, and structure
maps. Figure~\ref{fig:bar_lens_freq} shows the bar and lens\footnote{\added{We do not
    distinguish between lenses and ovals in spiral galaxies.}} fraction as a function of Hubble
type. The bar fraction among S0s is significantly lower than that of spirals
(\deleted{$59\pm3\%$}\added{$56\pm3\%$}), but their lens fraction is much larger than that of
spirals (\deleted{$7\pm2\%$}\added{$8\pm2\%$}). Interestingly, once lenses are taken into
consideration, the statistics of S0s become compatible with those of early-type spirals (Sa--Sb). It
is worth noting that the bar+lens fraction of S0s is considerably larger than that of late-type
spirals ($\sim60\%$).

The bar fraction of S0s quoted here is lower than that given by
\citeauthor{2009ApJ+Laurikainen}~(\citeyear{2009ApJ+Laurikainen}; $46\pm5\%$) or
\citeauthor{2017ApJ+Li}~(\citeyear{2017ApJ+Li}; $\sim50\%$), the latter based on CGS itself. We
attribute these differences to our conservative identification of barred galaxies---only definitive
cases are classified as barred. For example, the buckled bars in \citet{2017ApJ+Li} might be
classified as lenses in this study, because they are less flattened compared with those that are not
buckled. On the other hand, possibly missing weak bars would be included among the statistics for
lenses, such that the bar+lens fraction should be a more robust indicator of large-scale
nonaxisymmetric structure.

\subsection{Environments}
\label{sec:environments-s0s}

Most (83\%) of our S0s reside in groups. In the context of the entire CGS sample, the frequencies of
S0s in the field, groups, and \deleted{clusters}\added{Fornax} are $3\pm1\%$, $13\pm2\%$, and
$15\pm6\%$, respectively.  \deleted{Therefore, to the magnitude limit of the survey
  ($B_{T}\leq12.9\,\mathrm{mag}$), S0s in groups are as prevalent as in clusters.} Note that the
\added{frequency of S0s in Fornax is much lower than those in nearby galaxy clusters ($\sim50\%$;
  \citealp{1980ApJ+Dressler}), due to the fact that Fornax is not a representative
  cluster.}\deleted{only galaxy cluster covered by CGS is Fornax, which is not representative.  This
  may be responsible for the low frequency of S0s in the cluster environment in CGS.}

\begin{figure*}
\epsscale{1.16}
\plotone{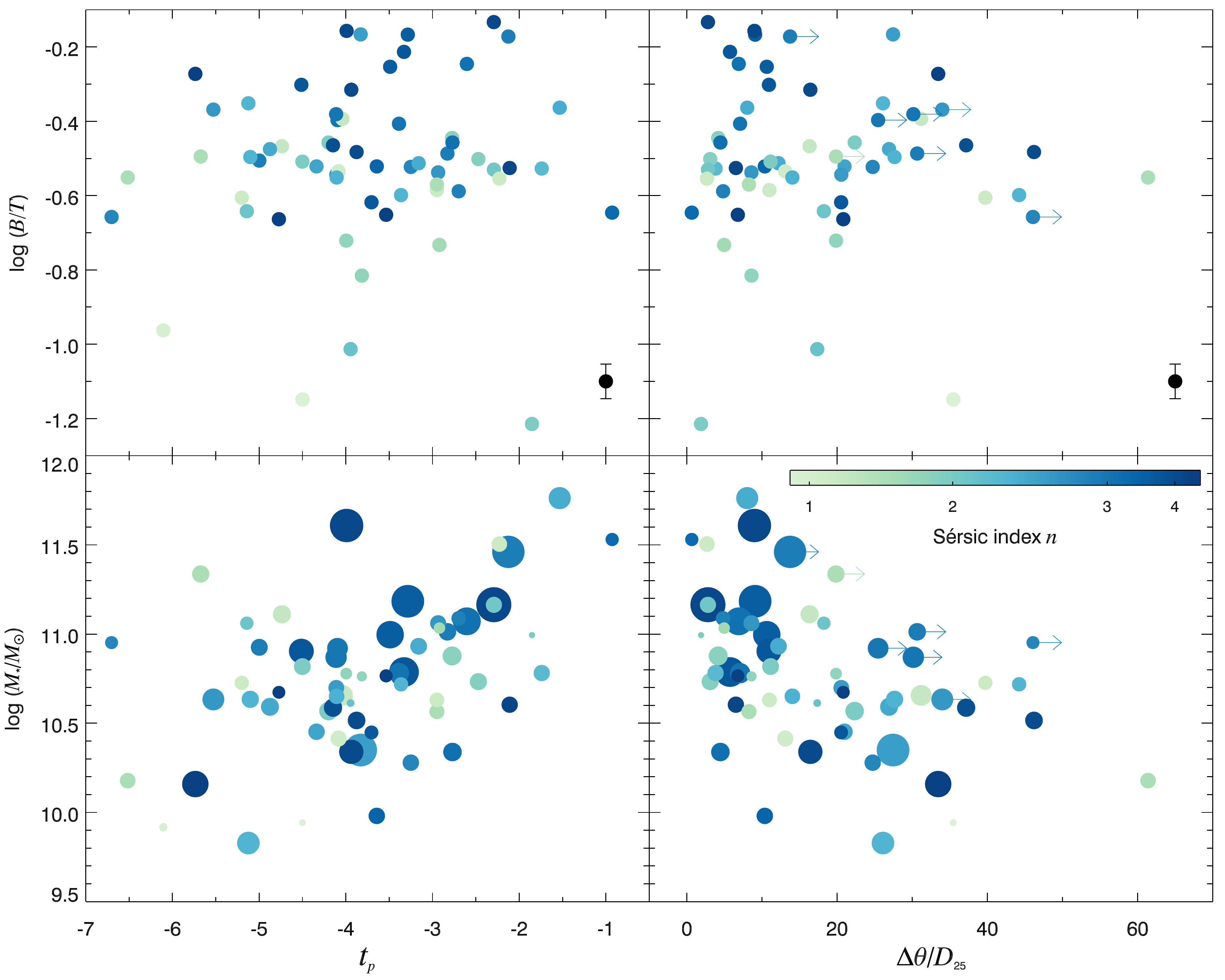}
\caption{Dependence of (top) $B/T$ and (bottom) $M_{\star}$ on two measures of local environment
  (see \citealp{2011ApJS+Ho} for definitions): (left) the tidal parameter $t_p$; (right) projected
  angular separation to the nearest bright neighbor $\Delta\theta/D_{25}$. In the right panels,
  \added{symbols with right arrows indicate that their $\Delta\theta/D_{25}$ are lower limits;} two
  galaxies are omitted to ensure better choice of $x$-axis range. The black
  \deleted{point}\added{filled circle} with error bars on the bottom-right corner of the upper
  panels illustrates the mean uncertainty in $\log{} (B/T)$. In all panels, the symbol color is
  assigned according to their bulge $n$, with darker color meaning larger $n$, as indicated on the
  color bar in the bottom right panel. In the bottom two panels, larger symbols indicate larger
  $B/T$. \label{fig:bul_env}}
\end{figure*}

It is almost unavoidable to think about S0s in the context of environments. As mentioned in
Section~\ref{sec:introduction}, many studies suggest that environmental processes play an important
role in the course of S0 production.  \citet{2011ApJS+Ho} provide two quantitative indicators of
local environment for CGS: (1) the tidal parameter
\begin{equation}
t_p \equiv \log \left\{ \sum_{i} \frac{M_i}{M_0}
\left(\frac{R_0}{D_i}\right)^{3} \right\} ,
\end{equation}
where $M_0$ and $R_0$ are the mass and size of the galaxy in question, and $M_i$ and $D_i$ are the
mass and projected separation of neighbor $i$; (2) the projected angular separation,
$\Delta \theta$, in units of the optical $B$-band diameter $D_{25}$, to the nearest neighboring
galaxy having an apparent magnitude brighter than $B_T\,+\,1.5$ mag. Both parameters are calculated
for neighbors within a radius of 750 kpc and with systemic velocity within
$\upsilon_{h}\pm500$~km~s$^{-1}$. The top row of Figure~\ref{fig:bul_env} shows the dependence of
$B/T$ and $n$ on these measures of local environment.  Although the correlations are statistically
weak, there is a mild tendency for S0s with large $B/T$ or $n$ to reside preferentially in denser
environments. Namely, for S0s with $B/T>0.5$, 89\% of them have $t_{p}>-4$; for S0s with $n>4$, 75\%
of them have $t_{p}>-4$. Similarly, 78\% of S0s with $B/T>0.5$ have $\Delta\theta/D_{25}<20$, and
63\% of S0s with $n>4$ have $\Delta\theta/D_{25}<20$.

The interpretation of these trends, however, is complicated by the known strong dependence of
stellar mass on environment. The most massive galaxies preferentially reside in dense environments
\citep{2013MNRAS+Calvi}, as confirmed in the bottom two panels of Figure~\ref{fig:bul_env}. The
stellar masses $M_{\star}$ were derived from total $K_s$ magnitudes from \citet{2011ApJS+Ho} and
mass-to-light ratios following Equation~(9) in \citet{2013ARA&A+Kormendy}, utilizing $B-V$ colors
from CGS \citep{2011ApJS+Li}. We color-code the symbols according to their bulge $n$ and adjust the
symbol size based on $B/T$.  At a fixed narrow range in $M_{\star}$ (a horizontal cut), we find no
compelling evidence for systematic variation of $n$ or $B/T$ with \deleted{environment}\added{the
  environmental indicators}. Therefore, we conclude that \added{after isolating mass effects,} the
\deleted{apparent}\added{aforementioned} trends between bulge properties and \deleted{environments
  are driven mostly or entirely by mass effects}\added{environmental indicators vanish}.

\section{Implications}

\subsection{The Surprising Homogeneity of Bulges in S0s}

As with \citet{2005MNRAS+Laurikainen}, we find that the bulges of S0s exhibit a broad distribution
of $B/T$ and $n$, in qualitative agreement with the notion that S0 galaxies form a parallel sequence
with spiral galaxies.  Are S0s simply defunct spirals?  Closer inspection exposes some tensions with
this proposition. In particular, while the more massive bulges of bulge-dominated S0s do bear a
close resemblance to the bulges of early-type spirals, the characteristically smaller bulges of
lower mass S0s do \textit{not} obey the same scaling relations as the bulges of late-type spirals
\citep{2010MNRAS+Laurikainen}. The bulges of late-type spirals---often of the pseudobulge
variety---appear as low-$\mu_e$ outliers in the Kormendy relation, whereas the comparably sized
bulges of S0s do not (see their Figure~2). If late-type spirals were to evolve to become late-type
(low-$B/T$) S0s, they would have to make their bulges denser (higher $\mu_e$) and more compact
(smaller $r_e$) to conform to the Kormendy relation of S0 galaxies. No simple fading scenario can
transform the population of small (pseudo) bulges in late-type spirals to the bulges currently
residing in low-mass S0s. Our present analysis arrives at a similar conclusion. While we have not
yet completed the decomposition of the full parent sample of CGS galaxies to perform a proper
comparison between the bulges of S0s and the bulges of spirals analyzed in the same manner, the
\deleted{uniformly tight}\added{uniform behavior of the} Kormendy relation shown in Figure~\ref{fig:FP}a
implies that the bulges of S0s constitute a homogeneous population, one that leaves little room for
a second population. This conclusion stands in sharp contrast to previous work that supports the
existence of pseudobulges in S0s \citep{2005MNRAS+Laurikainen,2006AJ+Laurikainen,
  2007MNRAS+Laurikainen}. They used photometric ($n\la 2$) and kinematic features (rotation support)
to identify pseudobulges. If we adopt a similar approach to identify pseudobulges using the
criterion $n\leq 2$, we would designate $\sim 31\%$ as pseudobulges.
\citet{2007ApJ+Barway,2009MNRAS+Barway} and \citet{2013ApJ+Vaghmare} classify pseudobulges based on
distributions of bulge parameters in various scaling relations (e.g., the bulge-disk correlation,
the Kormendy relation, the photometric plane). They conclude that pseudobulges are prevalent in
galaxies with absolute $K_s$-band magnitudes fainter than $-24.5$. As the majority (76\%) of the CGS
S0s lie below this critical luminosity threshold, many should qualify as pseudobulge hosts. However,
no such population of pseudobulges stands out in our scaling relations.

Part of the difficulty lies in the fact that no single observational criterion uniquely defines the
pseudobulge phenomenon (see Supplemental Information in \citealp{2013ARA&A+Kormendy}).
\added{\sersic{} indices \citep[e.g.,][]{2008AJ+Fisher}, morphologies
  \citep[e.g.,][]{2008AJ+Fisher}, the Kormendy relation \citep[e.g.][]{2009MNRAS+Gadotti}, and
  kinematics \citep[e.g.,][]{1982ApJ+Kormendy} are indicators commonly used for bulge
  classification. \citet{2017A&A+Neumann} suggest that the combination of the Kormendy relation and
  the concentration index gives the most robust classification of bulges types.} In the current
literature, the most widely adopted criterion for recognizing pseudobulges is a low \sersic{} index
($n<2$; \citealp{2008AJ+Fisher}), but for our present sample, despite our enormous effort to
quantify measurement uncertainties \citep{2017ApJ+Gao}, the \sersic{} index does not appear to
provide much discriminating power. Nearly one-third of our sample have $n<2$, and yet none of these
objects, which would ordinarily be regarded as pseudobulges, stands out in any noticeable manner in
either the Kormendy or the fundamental plane relations (Figures~\ref{fig:FP} and
\ref{fig:FPres}). Not all galaxies with small $B/T$ have pseudobulges \citep{2013ARA&A+Kormendy},
but essentially all pseudobulges have $B/T\la 1/3$ \citep{2009MNRAS+Gadotti,2016ASSL+Kormendy}. This
value coincides with the mean $B/T$ of our sample, and we expect that at least some of these
disk-dominated galaxies to contain pseudobulges. This echoes similar conclusions reached by
\citet{2017A&A+Costantin} for a small sample of late-type spirals. As discussed in
Section~\ref{sec:nonun-unif-s0}, the subset of bulges with disky morphology (blue symbols in
Figure~\ref{fig:FP}) is also unremarkable from the rest of the sample lacking disky features. Closer
examination of these 12 pseudobulge candidates reveals that in all but three (NGC~1386, NGC~4802,
NGC~4984) the disky features are embedded in a structure very much resembling a classical bulge once
the nuclear substructure is properly considered in the image decomposition. A good example is
NGC~1326. \citet{2017ApJ+Gao} show that properly masking the nuclear ring in this galaxy boosts its
bulge \sersic{} index from 1.29 to 2.01. Thus, morphological features such as disks do not reliably
signify the overall photometric structure \added{or the star formation activities} of the bulge (see
also \citealp{2009ApJ+Fisher}). \added{To summarize the above discussion: our application of the
  Kormendy relation and the fundamental plane is more robust than using bulge $n$ or morphology to
  recognize bulge dichotomy.}

\subsection{Formation Mechanisms of S0s}

If S0s arise from the simple fading of spirals, the abundant pseudobulges of late-type spirals should
have been preserved among lower mass S0s of relatively moderate to low $B/T$. This population of
pseudobulges appears conspicuously missing in S0s. To reconcile with the scenario that S0s derive
from faded spirals, some additional processes must operate to alter the structure of their bulges so
that they become denser (higher $\mu_e$) and more compact (smaller $r_e$), in order to conform to
the fundamental plane relations (Figure~\ref{fig:FP}). Candidate physical mechanisms include tidal
interactions, minor mergers, and galaxy harassment, which may contribute to bulge growth via
non-secular processes (e.g., nuclear starbursts and dissipationless mergers with satellites) that,
at the same time, can erase the disky origin of the bulges. Whether or not this can actually be
realized needs to be verified with numerical simulations.

However, if spirals transform to S0s via this pathway, we would expect the bulge properties of S0s
to show some dependence on \added{the relevant} \deleted{environment}\added{environmental
  indicators}. Figure~\ref{fig:bul_env} argues against this possibility, \deleted{and instead
  suggests that bulge buildup in S0s more likely involved mass-dependent processes}\added{especially
  after isolating effects due to mass}. An alternative, more radical possibility is that
pseudobulges simply never existed in \added{these mostly group} S0s, that they host exclusively
classical bulges. Most S0s are not faded \added{late-type} spirals, and their bulges formed from a
fundamentally different channel, at an earlier epoch. Bulges in early-type disk galaxies (such as
S0s) may have been in place since $z\approx 2$ \citep{2015ApJ+Graham}, when clumps formed out of
turbulent disks migrate inward and merge to contribute to bulge growth
\citep[e.g.,][]{2008ApJ+Genzel,2016ASSL+Bournaud, 2017ApJ+Tadaki}, although the contribution from
mergers to bulge buildup may still have been substantial
\citep[e.g,][]{1977egsp+Tmoore,2010ApJ+Hopkins}. Disky features created via later secular evolution
would not be able to alter significantly the physical properties of the preexisting bulges.

This scenario of bulge formation and galaxy evolution is still incomplete. Outstanding questions
remain. How did the early-formed compact bulges of S0s grow their disks, and how did their classical
bulges remain so pristine against subsequent secular evolution of the disks? How were the thin and
thick disks observed at $z\approx 0$ assembled? Are the ubiquitous thick disks \added{found in
  present-day S0s} descendants of $z \approx 2$ clumpy disks \citep[e.g.,][]{2006ApJ+Elmegreen,
  2009ApJ+Bournaud}? If bulges can naturally arise out of the clumpy disks without mergers, why did
present-day late-type spirals and pure disk galaxies \citep[e.g.,][]{2004ARA&A+Kormendy,
  2016ApJ+Sachdeva} fail to assemble noticeable classical bulges at that epoch? These issues deserve
further investigation but are certainly beyond the scope of our study.

To summarize this and the preceding section: our photometric study of S0 bulges casts doubt on the
idea that present-day spirals are the progenitors of S0 galaxies. Our argument is based mainly on
the absence of pseudobulges in S0s, as judged by the Kormendy and fundamental plane relations of
their bulges. The evidence is most striking for less massive, more disk-dominated systems, where
pseudobulges usually dominate, but we cannot preclude the possibility that the phenomenon extends to
the entire S0 class. \citet{2015A&A+Querejeta2} have reached similar conclusions on the basis of
stellar angular momentum and concentration measurements from the CALIFA survey.

\subsection{Lenses May Be Dissolved Bars}
\label{sec:lenses-may-be}

The bar fraction of the CGS S0s is significantly lower than that of spirals. However, combining the
statistics of bars and lenses in S0s greatly reduces the difference. The bar+lens fraction of S0s
(81\%) agrees well with the bar fraction of spirals of early to intermediate type, but is
considerably larger than that of late-type spirals ($\sim60\%$). This independently supports our
suggestion that late-type spirals are not plausible progenitors of S0s.

The dearth of bars in S0s can be understood as the consequence of bar-driven secular evolution,
during which bars self-destroy after the build-up of a sufficiently large central mass condensation.
Lenses may be the evolved remnants of bars.
In accordance with this expectation, observations show that bars in massive early-type disks tend
to be more buckled \citep{2017ApJ+Li} and exhibit weaker bar torques \citep{2007MNRAS+Laurikainen}.
If the assumption that lenses in S0s represent defunct bars is true, then major mergers could not
have played a major role in their recent evolution, lest the slow, secular processes of bar
evolution---an inherently disk phenomenon---be disrupted. Of course, the very existence of a
prominent large-scale disk in S0s, too, precludes the possibility of much recent dynamical violence.
We acknowledge the possibility that stellar disks can be rebuilt from gas-rich mergers
\citep[e.g.,][]{2015MNRAS+Wang,2016ApJ+Athanassoula,2017MNRAS+Sparre}, but even if this were true, the merger
event does not directly participate in the transformation of spirals to S0s but rather predates it.

The evolutionary link between bars and lenses, it should be stressed, is in no way settled.
\citet{1983IAUS+Athanassoula} suggested that lenses can arise from disk instabilities in a similar
way as bars do, with the initial velocity dispersion of the disk being a main determinant in whether
a bar or a lens appears. \citet{1983IAUS+Bosma} proposed that lenses form as a result of truncation of
star formation. Stable bars generally form easily in numerical simulations, stable, but few studies
can track the entire process of a bar evolving to a lens (but see \citealp{2002A&A+Bournaud}). Most
of observational evidence relating bars to lenses \citep[e.g.,][]{1979ApJ+Kormendy,
  2009ApJ+Laurikainen,2018MNRAS+Kruk} is circumstantial, including ours. More effort is needed to
complete this picture.

\subsection{Preprocessing of S0s in Groups}
\label{sec:prepr-s0s-groups}

\deleted{In this study, the frequency of S0s in groups is as high as that in clusters. The
  statistics based on CGS, however, may not be representative. The fraction of S0s in local galaxy
  clusters is $\sim0.5$ \citep{1980ApJ+Dressler}, much larger than our estimate.}\added{It is
  interesting to note that S0s are not exclusively cluster phenomena---most of the CGS S0s reside
  in groups. Unfortunately, with only Fornax as a member, CGS does not include representative galaxy
  clusters to enable proper comparison of S0 frequencies in various environments.} Nevertheless,
studies apart from ours \deleted{also} report S0 fractions in groups as large as those in clusters,
both at $z\approx0$ \citep{2003MNRAS+Helsdon,2009ApJ+Wilman} and at $z\approx0.4$
\citep{2009ApJ+Wilman}. S0 fractions evolve more rapidly in groups \citep{2010ApJ+Just}, and it is
likely that a substantial fraction of S0s completed their morphological transformation before they
enter clusters.

What physical mechanisms are responsible for producing S0s in groups?  A major difference between
group environment and cluster environment is that galaxies in groups are more likely to merge and
are less influenced by the intergalactic medium. The popularly discussed process of ram-pressure
stripping should still operate in groups to some extent, but whether the intragroup medium is dense
enough to strip significant amounts of disk gas is in question \citep[e.g.,][]{2006MNRAS+Rasmussen}.
Transformation of spirals into S0s by major mergers seems unlikely, as discussed in
Section~\ref{sec:lenses-may-be}. By contrast, minor mergers may be a plausible candidate to
accelerate gas consumption and feed central black hole accretion \citep{2014MNRAS+Kaviraj}, either
by triggering disk instabilities or by accretion of counter-rotating gas that facilitates gas
inflow. However, truncation of gas resupply is needed to keep the galaxy quenched. Ram-pressure
stripping of the hot halo gas has been shown to be possible even in small groups, thus providing a
possible explanation \citep{2009MNRAS+Bekki}.

Although Figure~\ref{fig:bul_env} suggests that environmental processes do not play a significant
role in building the bulges of S0 galaxies, they might affect their disks
\citep[e.g.,][]{1990AJ+Cayatte}. Galaxy disks come in three types, depending on the shape of their
radial profile \citep{2006A&A+Pohlen,2008AJ+Erwin}: Type I disks have a single, pure exponential
profile; Type II disks have a main exponential profile that downturns to a steeper profile at large
radii; Type III disks show an upturn at outer radii. \citet{2011ApJS+Li} found that the fraction of
Type~\Rmnum{1} disk profiles is nearly constant across the Hubble sequence, while the disks of
early-type galaxies have a deficit of Type~\Rmnum{2} profiles but an excess of the Type~\Rmnum{3}
variety compared with late-type systems.  The disk profiles of S0s depend on environment: Virgo
cluster S0s lack Type~\Rmnum{2} disks but have more Type~\Rmnum{1} disks compared with field S0s
\citep{2012ApJ+Erwin}. These trends imply that the formation of S0s may involve transformation of
their disks through various environmental processes \citep[e.g.,][]{2017MNRAS+Clarke}. We will
address this issue in a companion paper (H. Gao et al. 2018, in preparation).

While this section mainly focuses on group-related physical processes, we do not intend to exclude
the possibility that S0s can be produced in other environments. Most of the above discussion applies
to clusters as well. For instance, stronger ram-pressure stripping and high-speed encounters make
galaxies more vulnerable to tidal interactions.  Minor mergers and tidal interactions, albeit less
efficient in low-density environments, can also account for the production of the minority of S0s
observed in the field.

\section{Summary}
\label{sec:summary}

To shed light on the formation mechanism of S0 galaxies, we present a set of homogeneous
measurements of the optical structural parameters of the bulges of a well-defined sample of 62 S0
galaxies selected from CGS, based on high-quality, detailed two-dimensional image decomposition. We
also quantify their frequency of bars and lenses, as well as their environments. The bulges of S0
galaxies show a broad distribution of bulge-to-disk ratios ($B/T\approx 0.1-0.7$;
$\mathrm{mean}=0.34\pm0.15$) and \sersic{} indices ($n \approx 0.4-5.0$;
$\mathrm{mean}=2.62\pm1.02$), qualitatively consistent with the notion that S0s define a parallel
sequence similar to spiral galaxies. However, unlike \added{late-type} spirals, the S0s in our
sample show little evidence for pseudobulges. The bulges define a uniform, homogeneous sequence on
the Kormendy and fundamental plane relations. S0s have fewer bars but more lenses than spirals,
suggesting that lenses may be dissolved bars. If so, this limits the role that major mergers could
have played in the recent evolution of S0s. \deleted{CGS S0s reside in groups as frequently as in
  clusters, but}\added{S0 production does not occur exclusively in clusters, as most of the CGS S0s
  are found outside clusters.} \deleted{The environment does not seem to have much direct effect on
  their bulges.} \deleted{Many}\added{Some, if not many,} S0s likely were preprocessed and completed
their transformation in groups. \added{The structural properties of S0 bulges do not correlate with
  the tidal parameter or the projected angular separation to the nearest bright neighbor, especially
  after isolating the effect of stellar mass. Thus, they are not likely transformed from
  pseudobulges by environmental processes such as tidal interaction, minor mergers, and galaxy
  harassment.} \deleted{The bulge component, however, seems to have been in place since early on.}
\added{These facts lead us to suggest that the bulges of S0s seem to have been in place since early
  on and that they formed in an intrinsically different manner compared with the bulges of late-type
  spirals. As S0 bulges were not pseudobulges then and now, we conclude that late-type spiral
  galaxies are not plausible progenitors of S0s.}

\acknowledgments

\added{We thank Robert Kennicutt, Yingjie Peng, Chengpeng Zhang, Stijn Wuyts, and Minjin Kim for useful
  discussions. An anonymous referee provided helpful comments that improved the manuscript.} This
work was supported by the National Key R\&D Program of China (2016YFA0400702) and the National
Science Foundation of China (11473002, 11721303). ZYL is supported by the Youth Innovation Promotion
Association, Chinese Academy of Sciences. His LAMOST Fellowship is supported by Special Funding for
Advanced Users, budgeted and administered by the Center for Astronomical Mega-Science, Chinese
Academy of Sciences (CAMS). This research has made use of the NASA/IPAC Extragalactic Database (NED)
which is operated by the Jet Propulsion Laboratory, California Institute of Technology, under
contract with the National Aeronautics and Space Administration.

\appendix

\section{Notes on Individual Galaxies}
\label{sec:notes-indiv-galax}

\textbf{ESO 221-G026}: The galaxy is classified as an elliptical in both HyperLeda and RC3, but
\citet{2013ApJ+Huang} discovered some substructures in it. The galaxy is likely to be an edge-on
system with a thin and a thick disk, which we provisionally denote as a lens and a disk in its
decomposition (see Figure~1.1). Since its bar/lens identification is not reliable (flagged as ``?''
in Table~\ref{tab:bul_param}), we do not take it into account when calculating bar/lens fraction.

\textbf{ESO 442-G026}: The galaxy is likely to be an edge-on system with a thin and a thick disk,
which we provisionally denote as a lens and a disk in its decomposition (see Figure~1.2). Since its
bar/lens identification is not reliable (flagged as ``?'' in Table~\ref{tab:bul_param}), we do not
take it into account when calculating bar/lens fraction.

\textbf{ESO 507-G025}: The galaxy is classified as an elliptical in HyperLeda but as an S0 in
RC3. We recognize a blue and dusty region around the galaxy center ($\sim30\arcsec$), and model it
as an extra disk component. The dust lanes are masked during the fitting.

\textbf{IC 2006}: The galaxy is classified as an elliptical in HyperLeda but as an S0 in RC3, and
\citet{2013ApJ+Huang} discovered some substructures in it. It has a nuclear lens and an inner lens.

\textbf{IC 2035}: In addition to an extremely small bulge, the galaxy hosts a short bar, two lenses,
and an underlying disk that exhibits different orientation. The inner lens is difficult to model
unless the outer lens is modeled simultaneously.

\textbf{IC 4329}: The galaxy is weakly barred. There is no disk break associated with the weak
bar. We need to fix some parameters of the bar component to ensure a reasonable fitting.

\textbf{IC 4991}: A ring-like pattern shows up on the residual image.  As we are not able to
identify a realistic ring structure and are unsure about its physical nature, we attribute this
pattern to artifacts and do not model it.

\textbf{IC 5267}: The galaxy has an inner disk whose surface brightness profile is reminiscent of a
lens. An outer ring is visible on the residual image. The dust lanes across the bulge is masked
during the fitting.

\textbf{NGC 254}: The galaxy has an inner lens and an outer ring. Inside $\sim5\arcsec$, we find
fine structures indicative of the presence of a nuclear ring and a nuclear bar. This galaxy was used
in \citet{2017ApJ+Gao} to illustrate that the outer lenses/rings can be ignored for the
purposes of bulge decomposition. Here we present the full details of its decomposition, with the
outer ring included in the model.

\textbf{NGC 584}: The galaxy is classified as an elliptical in both HyperLeda and RC3, but is
recognized as an S0 in \citet{2013ApJ+Huang}. It has a nuclear lens and an inner lens.

\textbf{NGC 936}: The galaxy has a bar that is enclosed by an inner ring. Its structural layout is
similar to NGC~1533.

\textbf{NGC 1201}: The galaxy contains an inner lens and an outer lens. But unlike normal cases with
two lenses of different sizes, the inner lens fills the outer lens in one dimension in this
case. Therefore, we also model the outer lens to avoid potential bias of bulge parameters. A
possible outer ring is visible on the residual image. There is a nuclear bar with a size of
$\sim5\arcsec$ and a PA $\approx$ $10\arcdeg$.

\textbf{NGC 1302}: This is a barred galaxy with an inner ring and an outer ring.  This galaxy was
used in \citet{2017ApJ+Gao} to illustrate that the outer lenses/rings can be ignored for the
purposes of bulge decomposition. Here we present the full details of its decomposition, with the
outer ring included in the model.
 
\textbf{NGC 1326}: The galaxy has a bar, a nuclear ring, an inner ring, and an outer ring. It is
part of the training sample presented in \citet{2017ApJ+Gao}. Here we show the decomposition results
that include the inner and outer ring, with the nuclear ring unmasked (Model3 in their
Table~8). Note that the uncertainties are different from those presented in their Table~8, since we
include the model-induced uncertainties in this study.

\textbf{NGC 1380}: The galaxy is likely to be an edge-on system with a thin and a thick disk, which
we provisionally denote as a lens and a disk (see Figure~1.15). Since its bar/lens
identification is not reliable (flagged as ``?'' in Table~\ref{tab:bul_param}), we do not take it
into account when calculating the bar/lens fraction. The ``lens'' component is not perfectly modeled by
the \sersic{} function. The dust lane running through the bulge is masked.

\textbf{NGC 1386}: The galaxy has a Type~\Rmnum{2} disk, on which spiral dust lanes are visible. We tried
to mask the majority of the dust lanes. The bulge is distinctly blue compared to the disk.

\textbf{NGC 1387}: The galaxy is barred and its disk is broken at the bar radius. The nuclear ring
is readily recognizable in the residual pattern and the color map.

\textbf{NGC 1400}: The galaxy is classified as an elliptical in HyperLeda but as an S0 in RC3. We
recognize a lens at $\sim20\arcsec$. The dust lanes are masked during the fitting.

\textbf{NGC 1411}: The galaxy has a nuclear lens and an inner lens. It is part of the training
sample presented in \citet{2017ApJ+Gao}. Here we show the decomposition results of the model that
includes the two lenses (Model3 in their Table~2). Note that the uncertainties are different from
those presented in their Table~2, since we include the model-induced uncertainties in this study.

\textbf{NGC 1527}: The galaxy has an inner lens and a weak outer lens. 

\textbf{NGC 1533}: The galaxy is barred and its disk is broken roughly at the bar radius. A
ring-like pattern in the central 10\arcsec{} implies the presence of a barlens---face-on version of
a boxy/peanut bulge. This galaxy is part of the training sample presented in \citet{2017ApJ+Gao}. Here
we show the decomposition results of Model2 in their Table~7. Note that the uncertainties are
different from those presented in their Table~7, since we include the model-induced uncertainties in
this study.

\textbf{NGC 1537}: The galaxy is classified as an elliptical in HyperLeda but as a weakly barred S0
in RC3. We recognize it as an S0 that has a nuclear lens and an inner lens.

\textbf{NGC 1543}: The galaxy has a nuclear bar, a large-scale bar, an inner lens/ring, and an outer
ring.

\textbf{NGC 1553}: The galaxy has a nuclear lens and an inner lens/ring. Thus, its model construction
is similar to that of NGC~1411.

\textbf{NGC 1574}: A bright foreground star is sitting on top of the galaxy disk. The galaxy has a
bar that is embedded in a lens. An outer ring is only visible on the residual image.

\textbf{NGC 1726}: There are dust lanes near the galaxy center, and we mask them during the fitting.

\textbf{NGC 1947}: There are many field stars throughout the image and dust lanes across the bulge.

\textbf{NGC 2217}: The galaxy has a bar, an inner and an outer ring. The model includes all these
features, because it is difficult to achieve reasonable fits to the bar and the inner ring without
the outer ring in the model. A nuclear ring with a size of $\sim10\arcsec$ is visible in the
residual pattern.

\textbf{NGC 2640}: The galaxy is weakly barred, and the disk is broken at the bar radius. A large
number of foreground stars are projected on top of the galaxy.

\textbf{NGC 2695}: The galaxy has an inner lens. 

\textbf{NGC 2698}: The galaxy is likely to be an edge-on system with a thin and a thick disk, which
we provisionally denote as a lens and a disk in its decomposition (see Figure~1.31). Since its
bar/lens identification is not reliable (flagged as ``?'' in Table~\ref{tab:bul_param}), we do not
take it into account when calculating bar/lens fraction.

\textbf{NGC 2781}: The galaxy has a nuclear ring, an inner lens/ring, and an outer ring. We do not
find any signature of a bar.

\textbf{NGC 2784}: The galaxy has an inner lens and an outer lens. It is part of the training sample
presented in \citet{2017ApJ+Gao}. Here we show the decomposition results of Model3 in their
Table~3. Note that the uncertainties are different from those presented in their Table~3, since we
include the model-induced uncertainties in this study.

\textbf{NGC 2983}: The galaxy is barred, and its disk is broken at the bar radius. Its model
construction is similar to NGC~1533.

\textbf{NGC 3056}: The galaxy has an inner lens/ring. The residual pattern seems to suggest the
presence of a nuclear lens, but we do not find significant signatures of substructures inside
$\sim20\arcsec$ from inspection of its image and isophotal analysis. So we do not pursue further
refinements of the model.

\textbf{NGC 3100}: The galaxy has two lenses, but their configuration is unlike that of a typical
inner-outer lens configuration.  One lens fills the other in one dimension; therefore, we model the
two lenses together. There are dust lanes near the bulge, which we mask during the fitting.

\textbf{NGC 3108}: This is an interesting case: a huge classical bulge is assembling a diffuse disk
around itself \citep{2008MNRAS+Hau}.

\textbf{NGC 3271}: Fortunately we do not need to deal with the disk break associated with the bar,
as the bulge is well embedded in the thick bar. The circular dust lane at the galaxy center is
masked during the fitting. We find fine structures that suggest the presence of a nuclear bar
roughly aligned with the large-scale bar.

\textbf{NGC 3892}: This barred galaxy has an inner ring and an outer ring. In addition, we need to
include a compact nucleus, which is modeled with a PSF component, or else the \sersic{} index of the
bulge would be unrealistically large.

\textbf{NGC 3904}: The galaxy is classified as an elliptical in both HyperLeda and RC3, but is
recognized as a possible S0 in \citet{2013ApJ+Huang}. It has two lenses, one filling the other in
one dimension. We model both lenses simultaneously.

\textbf{NGC 4024}: The galaxy is barred, and its disk break at the bar radius is weak. Its model
construction is similar to NGC~1533.

\textbf{NGC 4033}: The galaxy is classified as an elliptical in both HyperLeda and RC3, but is
recognized as a possible S0 in \citet{2013ApJ+Huang}. It has a nuclear lens.

\textbf{NGC 4373A}: The galaxy is likely to be an edge-on system with a thin and a thick disk, which
we provisionally denote as a lens and a disk in its decomposition (see Figure~1.43). Since its
bar/lens identification is not reliable (flagged as ``?'' in Table~\ref{tab:bul_param}), we do not
take it into account when calculating bar/lens fraction. The dust lane running through the bulge is
masked during the fitting.

\textbf{NGC 4546}: The galaxy is likely to be an edge-on system with a thin and a thick disk, which
we provisionally denote as a lens and a disk in its decomposition (see Figure~1.44). Since its
bar/lens identification is not reliable (flagged as ``?'' in Table~\ref{tab:bul_param}), we do not
take it into account when calculating bar/lens fraction.

\textbf{NGC 4684}: We attribute the lens-like structure with a size of $\sim20\arcsec$ as the
bulge. Otherwise, the galaxy would have $B/T = 0$. The compact nucleus is modeled as a PSF
component. The central dust lane is masked during the fitting.

\textbf{NGC 4697}: The galaxy is classified as an elliptical in both HyperLeda and RC3, but is
recognized as an S0 in \citet{2013ApJ+Huang}. The galaxy is likely to be an edge-on system with a
thin and a thick disk, which we provisionally denote as a lens and a disk in its decomposition (see
Figure~1.46). Since its bar/lens identification is not reliable (flagged as ``?'' in
Table~\ref{tab:bul_param}), we do not take it into account when calculating bar/lens fraction.

\textbf{NGC 4802}: The galaxy hosts a dusty but overall blue bulge, which is indicative of ongoing
star formation. In addition, we recognize a nuclear lens and an inner lens. We mask the dust lanes
around the bulge. The compact nucleus is modeled using a PSF component.

\textbf{NGC 4825}: The galaxy is classified as an elliptical in HyperLeda but as an S0 in RC3. The
central dust lane running through the bulge is masked during the fitting.

\textbf{NGC 4856}: The galaxy is relatively edge-on, but its bar is still readily recognized. Its
disk is broken at the bar radius.

\textbf{NGC 4984}: The galaxy has an inner lens and an outer ring. The bulge is distinctly blue
compared with the lens and the disk. This galaxy was used in \citet{2017ApJ+Gao} to illustrate that
the outer lenses/rings can be ignored for the purposes of bulge decomposition. Here we present the
full details of its decomposition, with the outer ring included in the model.

\textbf{NGC 5026}: The galaxy has a bar that is enclosed by an inner ring. An outer ring is visible
on the residual image.

\textbf{NGC 5266}: We mask the central circular dust lanes along the minor axis of the galaxy. 

\textbf{NGC 5333}: The galaxy has a nuclear lens and an inner lens.

\textbf{NGC 6673}: The galaxy is classified as an elliptical in HyperLeda but as an S0 in RC3, and
is recognized as a possible S0 in \citet{2013ApJ+Huang}. It has a nuclear lens and an inner lens.

\textbf{NGC 6684}: The galaxy has a bar, an inner ring, and an outer ring/lens. A nuclear bar
embedded in the bulge is roughly perpendicular to the large-scale bar.

\textbf{NGC 6893}: The galaxy has an inner lens and an outer lens. This galaxy was used in
\citet{2017ApJ+Gao} to illustrate that the outer lenses/rings can be ignored for the purposes of
bulge decomposition. Here we present the full details of its decomposition, with the outer
\deleted{ring}\added{lens} included in the model.

\textbf{NGC 6942}: The galaxy is barred and shows a disk break at $\sim50\arcsec$. Spiral patterns
are visible on the residual image, but they are quite weak and can be ignored.

\textbf{NGC 7049}: The galaxy has a lens. The circular dust lane around the bulge is masked during
the fitting.

\textbf{NGC 7079}: The galaxy has a bar and shows a disk break at $\sim40\arcsec$.

\textbf{NGC 7144}: The galaxy is classified as an elliptical in both HyperLeda and RC3, but is
recognized as an S0 in \citet{2013ApJ+Huang}. It has a nuclear lens and an inner lens.

\textbf{NGC 7192}: The galaxy is classified as an elliptical in both HyperLeda and RC3, but is
recognized as an S0 in \citet{2013ApJ+Huang}. It has a nuclear lens and an inner lens.

\textbf{NGC 7377}: The galaxy has a nuclear lens and an inner lens. The dust lanes are masked during
the fitting.

\bibliographystyle{aasjournal}
\bibliography{new.ms}

\end{CJK*}

\end{document}